\documentclass[aps,prd,tightenlines,showpacs,floatfix,preprint,
amssymb,byrevtex,nofootinbib]{revtex4}  %endfloats

\usepackage{amssymb,amsmath}
\usepackage{graphicx,bm}

\begin{document}

%%%%%%%%%%%%%%%%%%%%% Title %%%%%%%%%%%%%%%%%%%%%%

\title{Elliptic flow of deuterons in relativistic heavy-ion collisions}

%%%%%%%%%%%%%%%%%%%% Authors %%%%%%%%%%%%%%%%%%%%%
%%%%%%%%%%%%%%%%%%%% Addresses %%%%%%%%%%%%%%%%%%%%%

\author{Yongseok Oh}%
\email{yoh@comp.tamu.edu}

\affiliation{
Cyclotron Institute and Physics Department, Texas A\&M
University, College Station, TX 77843-3366, U.S.A.}

\author{Che Ming Ko}%
\email{ko@comp.tamu.edu}

\affiliation{
Cyclotron Institute and Physics Department, Texas A\&M
University, College Station, TX 77843-3366, U.S.A.}

\date{\today}

%%%%%%%%%%%%%%%%%%%% Abstract %%%%%%%%%%%%%%%%%%%%%

\begin{abstract}

Using a dynamical model based on the $NN \to d\pi$, $NNN \to dN$,
and $NN\pi \to d\pi$ reactions and measured proton and pion transverse
momentum spectra and elliptic flows, we study the production of
deuterons and their elliptic flow in heavy ion collisions at RHIC.
The results are compared with those from the coalescence model.
The deviation of deuteron elliptic flow from the constituent nucleon
number scaling expected from the coalescence model and the comparison
with the experimental data are discussed in connection to the allowed
nucleon phase space in these reactions.

\end{abstract}

\pacs{25.75.-q, 25.75.Dw, 25.75.Ld}

\maketitle

\section{Introduction}

The elliptic flow of particles in heavy ion collisions is a measure
of the strength of the second Fourier coefficient in the azimuthal
angle distribution of particle transverse momentum relative to the
reaction plane~\cite{Olli92,VZ94}. Significant information on the
properties of the hot dense matter formed during the initial stage
of heavy ion collisions has been obtained from the study of elliptic
flow~\cite{RR97,DLGP98,Lacey05}. In heavy ion collisions at the
Relativistic Heavy Ion Collider (RHIC)~\cite{STAR05c,PHENIX05a},
measured elliptic flow at low transverse momentum shows a mass
ordering, i.e., the strength of elliptic flow of identified hadrons
decreases with increasing hadron masses, and this has been well
described by ideal hydrodynamics~\cite{KSH00} as well as by the
transport model~\cite{LK01}. Furthermore, the elliptic flows of
identified hadrons, particularly at intermediate transverse momenta,
are seen to follow a constituent quark number scaling, i.e., the
dependence of hadron elliptic flows on hadron transverse momentum
becomes similar if both are divided by the number of constituent
quarks in a hadron. This scaling behavior of hadron elliptic flows
is consistent with the predictions of the quark coalescence model
for hadron production from produced quark-gluon plasma in
relativistic heavy ion
collisions~\cite{GKL03-GKL03b,HY02-HY03,FMNB03-FMNB03b,MV03,KCGK04}.

Recently, the elliptic flow of deuteron has also been measured at
RHIC~\cite{PHENIX06,PHENIX07,STAR07a}. These measurements show that
the deuteron elliptic flow seems to scale with its constituent
nucleon number, implying that the quark number scaling of elliptic
flows holds not only for hadrons but also for the deuteron. Since
deuteron production in heavy-ion collisions could be described by
the coalescence of protons and neutrons at freeze out~\cite{DHSZ91},
the observed nucleon number scaling of the deuteron elliptic flow is
thus not surprising. In the simplest coalescence
model~\cite{MV03,KCGK04} which involves only comoving particles, the
deuteron yield in momentum space is proportional to the product of
the proton and neutron densities at half the momentum of produced
deuteron, i.e., $\bm{p}_p=\bm{p}_n=\bm{p}_d/2$, and the deuteron
elliptic flow would satisfy exactly the nucleon number scaling and
thus the quark number scaling as well. This strong momentum
constraint is relaxed in the more general coalescence
model~\cite{DHSZ91} that takes into account the nucleon momentum
spread in the deuteron. As in the case of hadron production from
quark coalescence~\cite{GKL03-GKL03b}, the more general coalescence
model would lead to a small deviation of the scaled deuteron
elliptic flow from that of nucleons.

Although the momentum conservation is maintained in the coalescence
model, the energy conservation is not satisfied. Some doubt has thus
been raised on the applicability of the coalescence model,
especially in low transverse momentum region~\cite{McL07,RR07}.
Investigating the consequences arising from the energy conservation
condition is therefore required to test the coalescence model and to
understand the constituent number scaling of elliptic flows in low
transverse momentum region. In this work, this question will be
addressed by studying deuteron production and elliptic flow in
heavy-ion collisions based on a dynamical approach that satisfies
both energy and momentum conservations. Specifically, deuteron
production in the present study will be treated through $NN \to
d\pi$, $NNN \to dN$, and $NN\pi \to d\pi$ reactions. Since this
approach is based on physical scattering processes for deuteron
production, the energy-momentum conservation is always satisfied. By
comparing the physically allowed nucleon phase space for deuteron
production with that involved in the coalescence model, we can study
the consequences of the energy conservation condition. In a
realistic study, these reactions need to be included in the hadronic
stage of relativistic heavy ion collisions either via a microscopic
transport model~\cite{LKLZP04} or a schematic kinetic
model~\cite{CGKLL03} to take into account all possible deuteron
production and dissociation processes. Since the elliptic flow of
particles during the hadronic evolution does not change much as the
initial spatial asymmetry through which the elliptic flow is
generated has decreased significantly during the evolution of the
partonic stage~\cite{LK01}, we can therefore calculate the deuteron
yield and elliptic flow based on the deuteron production rate from
nucleons at freeze out.

This paper is organized as follows. In Section~\ref{coal}, we first
give a brief description of the coalescence model that will be used
to compare with the results from the dynamical model employed in the
present study for deuteron production. We then consider in
Section~\ref{two-body} and Section~\ref{three-body} deuteron
production from the two-body reaction $NN \to d\pi$ and three-body
reactions $NNN \to dN$ and $NN\pi \to d\pi$, respectively, with
their cross sections obtained from a hadronic model that is based on
empirically determined parameters~\cite{ADEK71}. In
Section~\ref{input}, nucleon and pion transverse momentum spectra
and elliptic flows, which are needed in both the coalescence model
and the dynamical model for studying deuteron production, are
discussed. Results on deuteron transverse momentum spectrum and
elliptic flow are then compared with available experimental data
from RHIC as well as those from the coalescence model in
Section~\ref{results}, where we further compare the nucleon phase
space involved in our dynamical model with that in the coalescence
model. Section~\ref{summary} contains the summary and discussions.
Details on the reaction amplitudes for both two-body and three-body
reactions are given in Appendixes~\ref{two} and~\ref{three},
respectively.

\section{Deuteron production in the coalescence model}
\label{coal}

In the coalescence model, the probability for the production of a
bound composite particle from a many-particle system is determined
by the overlap of the wave functions of coalescing particles with
the internal wave function of the composite
particle~\cite{Mattie95}. Assuming that nucleons are uniformly
distributed in space, the momentum spectrum of
deuterons formed from the coalescence of nucleons is given by
\begin{equation}
\frac{dN_d}{d^3{\bm p_d^{}}} =
\frac{3}{4} \frac{V}{(2\pi)^3} \int d^3{\bm p_1^{}} d^3{\bm p_2^{}}
f_p(\bm{p}_1^{}) f_n(\bm{p}_2^{})
|\Psi_d\bm{(}(\bm{p}_1'-\bm{p}_2')/2\bm{)}|^2
\delta^{(3)}(\bm{p}_1^{} + \bm{p}_2^{} - \bm{p}_d^{}).
\label{eq:coal}
\end{equation}
In the above, the factor $3/4$ is the probability for a proton and a
neutron to form a spin triplet state like that in a deuteron;
$f_p(\bm{p}_1^{})$ and $f_n(\bm{p}_2^{})$ are, respectively, the
proton and neutron momentum distributions in the hadronic matter of
volume $V$ at freeze out, and they are related to the proton number
$N_p$ and neutron number $N_n$ by $V \int \frac{d^3 {\bm
p}}{(2\pi)^3} f_{p,n}({\bm p})=N_{p,n}$; $\Psi_d$ is the deuteron
wave function in the momentum space with $\bm{p}_1'$ and $\bm{p}_2'$
denoting, respectively, the momenta of the proton and neutron in the
deuteron rest frame. In the more general case of non-uniform nucleon
distribution and/or with collective flow, additional spatial
integrals would appear in Eq.~(\ref{eq:coal}), and the deuteron
momentum space wave function is replaced by its Wigner function
which describes both the relative momentum and spatial distributions
of the two nucleons in a deuteron~\cite{CKL03,CKL03a}.

For the deuteron wave function, we use the one given by Hulth{\'e}n.
In the momentum space, it is given by
\begin{equation}
\Psi_d(k) =
\frac{\sqrt{(\alpha_d^{}+\beta_d^{})^3\alpha_d^{}\beta_d^{}}}
{\pi(\alpha_d^2 + k^2)(\beta_d^2 + k^2)}, \label{Hulthen}
\end{equation}
with $\alpha_d^{} = 0.23$ fm$^{-1}$ and $\beta_d^{} = 1.61$
fm$^{-1}$~\cite{CKL03a} and is normalized as $\int d^3 {\bm k}
|\Psi_d(k)|^2 =1$. It should be mentioned that this more general
coalescence model still does not respect the energy conservation
condition.

In the naive coalescence model, the deuteron wave function in the
momentum space is taken to be a delta function. In this case, the
transverse momenta $\bm p_{T,1}^{}$ and $\bm p_{T,2}^{}$ of the nucleons
forming the deuteron are restricted not only in magnitude but also
in direction so that $\bm{p}_{T,1}^{} = \bm{p}_{T,2}^{} =
\bm{p}_{T,d}^{}/2$ with $\bm{p}_{T,d}^{}$ being the deuteron
transverse momentum~\cite{KCGK04,MV03}. As a result, the scaling of
the nucleon elliptic flow $v_{2,N}$ and the deuteron elliptic flow
$v_{2,d}^{}$ according to their constituent nucleon numbers is exact,
i.e., $v_{2,d}^{}(p_T^{}) =
2v_{2,N}^{}(p_T^{}/2)$~\cite{KCGK04,yan}.

\section{\boldmath Deuteron production from two-body reactions}
\label{two-body}

For deuteron production from two-body nucleon-nucleon reactions, the
dominant reaction is $NN\to d\pi$ due to the small pion mass and the
strong pion-nucleon coupling. Indeed, experimental observations have
shown that the cross section for the $NN \to d\pi$ reaction is much
larger than the cross sections for deuteron production from other
two-body reactions~\cite{ADEK71}. In contrast to the coalescence
model, in which the deuteron is formed from a proton and a neutron
in the hadronic matter, not only the $pn$ reaction but also the $pp$
and $nn$ reactions can produce a deuteron in the $NN\to d\pi$
reaction as the pion, which has isospin one, can carry away the
mismatched isospin.

The production rate of deuterons with momentum $\bm p_d^{}$ from
the $NN \to d\pi$ reaction can be written as
\begin{eqnarray}
\frac{d\mathcal{R}_d}{d^3{\bm p_d^{}}} &=& \frac{1}{(2\pi)^3 2E_d} \int
\prod_{i=1}^2 \frac{d^3 \bm{p}_i^{}}{(2\pi)^3 2E_i}
f_N(\bm{p}_i^{}) \frac{d^3 \bm{p}_\pi^{}}{(2\pi)^3 2E_\pi}
[1+f_\pi(\bm p_\pi)] \nonumber\\
&& \mbox{} \times|\mathcal{M}(NN \to d \pi)|^2 (2\pi)^4\delta^{(4)}
\left( \sum_{i=1}^2 p_i^{}- p_\pi^{} - p_d^{}\right),
\label{eq:deut_2}
\end{eqnarray}
where $f_N(\bm p)$ and $f_\pi(\bm p_\pi^{})$
are nucleon and pion momentum distribution functions
with the latter normalized to the pion number in the hadronic
matter.
We have assumed here that the proton and neutron densities in the hadronic
matter are the same for simplicity.
The average of initial spin and isospin states
and sum over final spin and isospin states are included in the
squared transition amplitude $|\mathcal{M}(NN \to d\pi)|^2$. Details
on the amplitude for the $NN \to d\pi$ reaction are given in
Appendix~\ref{two}. Compared with the coalescence model,
Eq.~(\ref{eq:deut_2}) shows that both energy and momentum are always
conserved due to the produced pion. This allows nucleons with higher
momentum to produce a low-momentum deuteron since the pion can carry
off the excess energy of the two nucleons. As a result, there is no
\emph{a priori} restriction on the momenta such as $p_{T,d}^{}
\approx 2 p_{T,N}^{}$ as in the coalescence model. Instead, the
nucleon distribution and the energy-momentum conservation condition
determine the most plausible configurations of nucleon momentum for
producing the deuteron.

\section{\boldmath Deuteron production from three-body reactions}
\label{three-body}

Deuterons can also be produced from three-body reactions such as
$NNN\to dN$ and $NN\pi\to d\pi$. The reaction $NNN\to dN$ is known
to dominate deuteron production in low-energy heavy ion
collisions~\cite{CK86,DB91}. Because of the large abundance of pions
in the produced hadronic matter in heavy ion collisions at RHIC, the
reaction $NN\pi\to d\pi$ becomes also possible.

\subsection{\boldmath $NNN \to dN$ reaction}

For the $NNN \to dN$ reaction, the production rate for deuterons of
momentum $\bm p_d^{}$ is given by
\begin{eqnarray}
\frac{d\mathcal{R}_d}{d^3{\bm p_d^{}}} &=&
\frac{1}{(2\pi)^3 2E_d} \int \prod_{i=1}^3
\frac{d^3 \bm{p}_i^{}}{(2\pi)^3 2E_i} f_N(\bm p_i^{})
\frac{d^3 \bm{p}_N^{}}{(2\pi)^3 2E_N} \left[ 1-f_N(\bm p_N^{}) \right]
\nonumber\\
&& \mbox{}\times |\mathcal{M}(NNN \to dN)|^2 (2\pi)^4 \delta^{(4)}
\left(\sum_{i=1}^3p_i^{}- p_N^{} - p_d^{}\right). \label{eq:deut_3}
\end{eqnarray}

The presence of the third nucleon in the initial state leads to a
very different nucleon phase space in deuteron formation, as the
third (spectator) nucleon can ensure the energy-momentum
conservation. Since the nucleon mass is much larger than the pion
mass, the effect of the spectator nucleon is different from that of
the pion produced in the $NN \to d\pi$ reaction and thus could
affect the elliptic flow of produced deuteron differently. The
amplitude for the $NNN\to dN$ reaction can be obtained from its
inverse reaction $dN\to NNN$, which has been studied in
Refs.~\cite{DB91,DS93,GSTH06} based on the full $NN$ elastic
scattering amplitude. Instead of such a complete study, we
approximate the amplitude for the reaction $NNN\to dN$ by extending
the amplitude we derived in the previous section for the $NN\to
d\pi$ reaction to an off-shell pion that is absorbed by the third
spectator nucleon. Details on the modeling of the amplitude
$\mathcal{M}(NNN \to dN)$ are given in Appendix~\ref{three}.

\subsection{\boldmath $NN\pi \to d\pi$ reaction}

The same method introduced in the above for the $NNN\to dN$ reaction
can be used for the amplitude of the $NN \pi \to d\pi$ reaction,
i.e., the subprocess $NN\to d\pi$ is replaced by $NN\to d\rho$ with
the off-shell rho meson absorbed by the spectator pion. Details on
the modeling of the amplitude $\mathcal{M}(NN\pi\to d\pi)$ are also
given in Appendix~\ref{three}. In terms of the nucleon and
pion momentum distributions the production rate for deuterons with
momentum $\bm p_d^{}$ from this reaction is given by
\begin{eqnarray}
\frac{d\mathcal{R}_d}{d^3{\bm p_d}} &=& \frac{1}{(2\pi)^3 2E_d} \int
\prod_{i=1}^2 \frac{d^3 \bm{p}_i^{}}{(2\pi)^3 2E_i} f_N(\bm
p_i^{})\frac{d^3 \bm{p}_3^{}}{(2\pi)^3 2E_3} f_\pi(\bm p_3^{})
\frac{d^3 \bm{p}_\pi^{}}{(2\pi)^3 2E_\pi} [1+f_\pi(\bm p_\pi^{})]
\nonumber\\
&& \mbox{} \times|\mathcal{M}(NN\pi \to d\pi)|^2 (2\pi)^4
\delta^{(4)} \left(\sum_{i=1}^3p_i^{}- p_\pi^{} - p_d^{}\right).
\label{eq:deut_4}
\end{eqnarray}

\section{nucleon and pion transverse momentum distributions}
\label{input}

Both the coalescence model and the dynamical model use as input the
nucleon momentum distribution at freeze out.
Moreover, the pion momentum distribution at freeze
out is also needed in the dynamical model. For heavy ion
collisions at RHIC, both nucleon and pion rapidity distributions are
approximately uniform in the mid-rapidity region, i.e., $|y^{}|\le
0.5$, which we will assume in the present study. In this case, only
their transverse momentum distributions are needed. In the
following, we discuss how these distributions are obtained.

\subsection{Nucleon spectrum and elliptic flow}

For the nucleon $p_T^{}$ distribution $f_N(\bm{p}_T^{})$, we take it
to have the form
\begin{equation}
f_N (\bm{p}_T^{}) = \gamma_N^{} \exp\left( -m_T^{}/T_{\rm eff}
\right)[1 + v_{2,N}^{}(p_T^{}) \cos(2\phi)], \label{N_pt}
\end{equation}
where $m_T^{} = \sqrt{M_N^2 + p_T^2}$ with the nucleon mass $M_N$
and $\phi$ being the azimuthal angle of the nucleon transverse
momentum with respect to the reaction plane. We have introduced a
dimensionless parameter $\gamma_N^{}$ to compensate for the effects
caused by using the effective inverse slope parameter $T_{\rm eff}$
instead of the temperature of the system together with the radial
flow effect. The parameter $\gamma_N^{}$ also contains the density
of spin and isospin states as well as the possible effect due to
finite nucleon chemical potential.

\begin{figure}[t]
\centering
\includegraphics[width=0.6\hsize,angle=0]{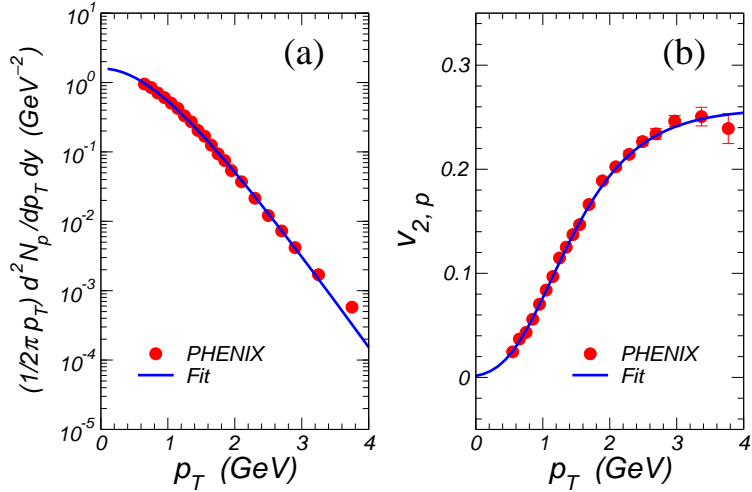}
\caption{(Color online) (a) $p_T^{}$ spectrum of protons. The solid
line is half of the parameterized nucleon distribution
[Eq.(\ref{N_pt})]. Experimental data are from Ref.~\cite{PHENIX03b}
for the minimum-bias proton spectrum (filled square). (b) Nucleon
differential elliptic flow $v_2^{}(p_T^{})$. The solid line is the
parameterized one [Eq.~(\ref{v2_N})] and filled squares are
experimental data from Ref.~\cite{PHENIX07}.} \label{fig:pt_spec}
\end{figure}

The inverse slope parameter in the nucleon $p_T^{}$ spectrum
[Eq.(\ref{N_pt})] is determined to be $T_{\rm eff}=295$ MeV by using
the experimental data of Ref.~\cite{PHENIX03b} for the minimum bias
proton $p_T^{}$ spectrum. Taking the effective volume of the
hadronic matter at freeze-out in minimum bias collisions to be
$V\approx 3,700~{\rm fm}^3$, which is about $1/3$ of the estimated
hadronic matter volume in central Au+Au collisions at
$\sqrt{s_{NN}^{}}=200$ GeV~\cite{CGKLL03}, we obtain $\gamma_N^{} =
0.021$ for the value of the parameter in Eq.~(\ref{N_pt}).%
\footnote{The difference of the proton $p_T^{}$ spectrum in minimum
bias and in central collisions shows that the volume of the hadronic
matter in the minimum bias collision is about $1/3$ of that in the
central collision when the data are parameterized using
Eq.~(\ref{N_pt}).} The comparison of our parameterized proton
$p_T^{}$ spectrum, which is half of the nucleon $p_T^{}$ spectrum,
with the experimental data is shown in panel (a) of
Fig.~\ref{fig:pt_spec}.

For the nucleon elliptic flow $v_{2,N}^{}$, we assume that it is the
same for both protons and neutrons and parameterize it in the form
of the Fisher-Tippet (or Gumbel) distribution
function~\cite{Gumbel}, i.e.,
\begin{equation}
v_{2,N}(p_T^{}) = \alpha_N^{} \exp \left\{-\exp \left[ \left(
\lambda_N^{} - p_T^{} \right) / \beta_N^{} \right] \right\}.
\label{v2_N}
\end{equation}
We use the experimental data of Ref.~\cite{PHENIX07} for the proton
$v_2^{}$, which were measured for 20--60\% centrality, to fix the
parameters. This gives $\alpha_N^{} = 0.258$, $\beta_N^{} =
0.683$~GeV, and $\lambda_N^{}=1.128$~GeV. This parametrization is
shown in panel (b) of Fig.~\ref{fig:pt_spec} together with the
experimental data of Ref.~\cite{PHENIX07}. Our parametrization gives
similar results in the measured $p_T^{}$ region as the
parametrization $a \tanh(b\, p_T^{} + c)$, which was introduced in
Ref.~\cite{GK04a} for quarks. But using the latter form would lead
to a negative $v_2^{}$ at small $p_T^{}$. Although there is no
experimental information on the nucleon $v_{2}^{}$ at very low
$p_T^{}$ region, hydrodynamic model studies~\cite{KHHH00,HKHRV01}
show that the nucleon $v_2^{}$ produced in heavy ion collisions at
RHIC has non-negative values~\cite{Lacey05}. Since the hydrodynamic
calculations are expected to reasonably describe $v_2^{}$ at low
$p_T^{}$ region, we employ the form of Eq.~(\ref{v2_N}) which gives
non-negative $v_{2,N}$ even for very small $p_T^{}$. We may also
take the proton $v_2$ to have the form $ \alpha' p_T^{} + \beta'
p_T^3$ for $ p_T^{} \leq p_T^0$ and $ a \tanh (bp_T^{} + c) $ for
$p_T^{} \geq p_T^0$, which gives a similar result as
Eq.~(\ref{v2_N}) if we choose $p_T^0 = 0.75$ GeV, $a=0.293$,
$b=0.460$ GeV$^{-1}$, $c=-0.175$, $\alpha'=0.0332$ GeV$^{-1}$, and
$\beta'=0.0579$ GeV$^{-3}$.

\subsection{Pion spectrum and elliptic flow}

\begin{figure}[t]
\centering
\includegraphics[width=0.6\hsize,angle=0]{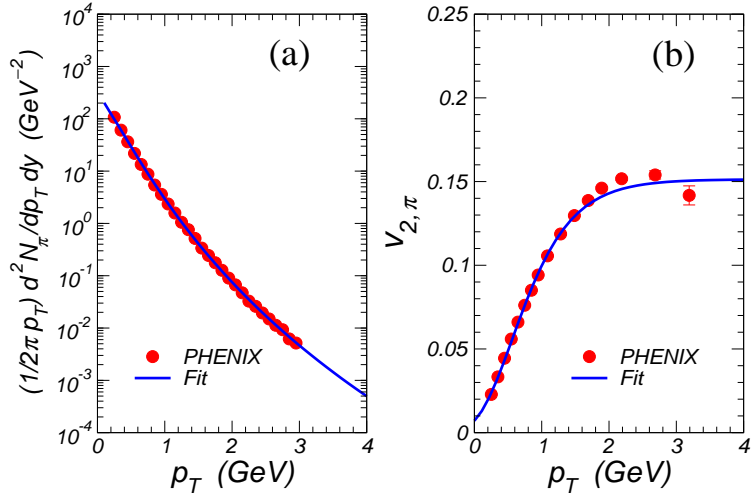}
\caption{(Color online) $p_T^{}$ spectrum (panel (a)) and $v_2^{}$
(panel (b)) of pions. Solid lines are parametrization of
Eqs.~(\ref{pi_pt}) and (\ref{pi_v2}). Experimental data are from
Refs.~\cite{PHENIX06,PHENIX03b}.}
\label{fig:pi_pt}
\end{figure}

For the pion $p_T^{}$ spectrum, the parametrization used for the
nucleon $p_T^{}$ spectrum [Eq.(\ref{N_pt})] does not work well as it
is affected by resonance decays~\cite{DESXX04,GK04a}. Instead, we
use the parametrization,
\begin{equation}
f_\pi(\bm{p}_T^{}) = \gamma_\pi^{} \left( 1 +
\frac{p_T^{}}{\alpha_\pi} \right)^{\beta_\pi} [1 +
v_{2,\pi}^{}(p_T^{}) \cos(2\phi)], \label{pi_pt}
\end{equation}
with $\alpha_\pi = 1.29$ GeV, $\beta_\pi = -12.0$, and $\gamma_\pi =
2.0$ determined from the experimental data of the PHENIX
Collaboration~\cite{PHENIX03b}. For the pion $v_2^{}$, it is
parameterized as in Eq.~(\ref{v2_N}),
\begin{equation}
v_{2,\pi}^{}(p_T^{}) = \alpha_\pi \exp \left\{ -\exp \left[ \left(
\lambda_\pi^{} - p_T^{} \right) / \beta_\pi \right] \right\},
\label{pi_v2}
\end{equation}
with $\alpha_\pi^{} = 0.184$, $\beta_\pi^{}=0.461$~GeV, and
$\lambda_\pi^{}=0.547$~GeV to fit the PHENIX experimental
data~\cite{PHENIX06}. The comparison of our parametrization with
experimental data is shown in Fig.~\ref{fig:pi_pt}.

\section{results}
\label{results}

With the above proton and pion $p_T^{}$ spectra and elliptic flows
as inputs, we calculate in this Section the deuteron $p_T^{}$
spectrum and $v_{2}^{}$ from the dynamical model and compare the
results with experimental data as well as those from the coalescence
model.

\subsection{Deuteron spectrum and elliptic flow from two-body reactions}

In terms of rapidity and transverse momentum,
Eq.~(\ref{eq:deut_2}) for the deuteron production rate can be
written as
\begin{eqnarray}
\frac{d^3 \mathcal{R}_d}{d^2{\bm p}_{T,d}^{}dy_d^{}} &=&
\frac{1}{8(2\pi)^8} \int \prod_{i=1}^2 dy_i^{} d{p}_{T,i}^{}
d\phi_i^{} \, p_{T,i}^{} f_N(\bm{p}_{T,i}^{}) [1+f_\pi({\bm
p_1}+{\bm p_2}-{\bm p_d})] \nonumber \\ && \mbox{} \times
|\mathcal{M}(NN \to d \pi)|^2 \delta\bm{(} (p_1^{}+p_2^{} -
p_d^{})^2 - M_\pi^2 \bm{)}. \label{eq:det2}
\end{eqnarray}
The deuteron $p_T^{}$ spectrum is then obtained by multiplying the
above rate by the volume of the hadronic matter at freeze out
$V\approx 3,700~{\rm fm}^3$ and the time interval $\Delta\tau\approx
4~{\rm fm}/c$. The latter value is chosen to reproduce reasonably
measured deuteron spectrum. It takes into account the fact that in a
more microscopic approach that follows the production and
dissociation of deuterons from the reaction $NN\to d\pi$ as well as
the deuteron elastic scattering, deuterons are expected to be
produced over a finite time interval from the expanding hadronic
matter formed in heavy ion collisions. To extract the elliptic flow
$v_{2}^{}$ of produced deuterons, we further express the result as
\begin{equation}
\frac{d^3 N_d}{p_{T,d}^{} d p_{T,d}^{}d\Phi_d dy_d^{}} =
f_d(p_{T,d}^{},y_d) [ 1 + 2 v_{2,d}^{}(p_{T,d}^{}) \cos(2 \Phi_d) ],
\end{equation}
where $\Phi_d$ is the azimuthal angle of produced deuteron and
$\phi_{1,2}^{}$ are those of the two nucleons relative to the
reaction plane. Because of the $\delta$-function which comes from
the energy-momentum conservation and the pion on-shell condition,
one of the nucleon azimuthal angles is constrained by kinematics.
This is different from the assumption in the naive coalescence model
that the direction of the nucleon momentum is constrained to $\phi_1
= \phi_2 = \Phi_d$.

\begin{figure}[t]
\centering
\includegraphics[width=0.6\hsize,angle=0]{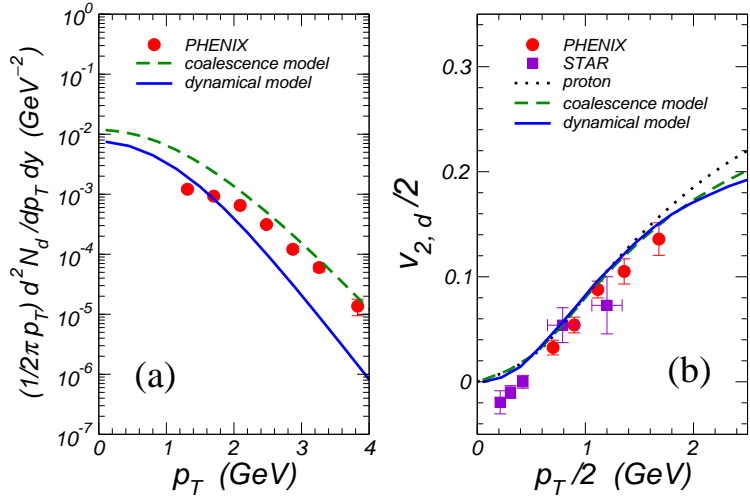}
\caption{(Color online) (a) $p_T^{}$ spectrum of deuterons.
Experimental data are from Ref.~\cite{PHENIX05} for the minimum-bias
collisions (filled circle). (b) Scaled deuteron elliptic flow
$v_2^{}$ as a function of $p_T^{}/2$ at $y_d^{}=0$. The dotted line
is the nucleon elliptic flow $v_{2,N}^{}$ of Eq.~(\ref{v2_N}).
Experimental data are from the PHENIX Collaboration~\cite{PHENIX07}
(filled circles) and the STAR Collaboration~\cite{STAR07a} (filled
squares). In both figures, the dashed and solid lines are results
from the coalescence model of Eq.~(\ref{eq:coal}) and the dynamical
model in the present work, respectively.} \label{fig:pt_spec
deuteron}
\end{figure}

The resulting deuteron $p_T^{}$ spectrum using the phenomenological
scattering amplitude given in Appendix~\ref{two} for the $NN \to
d\pi$ reaction is shown by the solid line in panel (a) of
Fig.~\ref{fig:pt_spec deuteron}. Compared with the measured $p_T^{}$
spectrum of deuterons shown by filled circles, ours is softer. This
result indicates that the radial flow effect on produced deuterons
is not fully taken into account in the dynamical model. Since the
cross sections for $\pi d$ elastic and inelastic scatterings are
comparable at pion kinetic energy of around $100$~MeV \cite{KDHM91},
deuterons produced from the dynamic model are expected to suffer
further scatterings and thus to approach thermal equilibrium with
pions. To properly treat the radial flow effect on produced deuteron
spectrum requires the inclusion of the processes considered in
present study in a non-equilibrium transport model, such as the ART
model~\cite{art}. As we are mainly interested in studying the effect
of energy conservation on the deuteron elliptic flow, we will leave
such more realistic approach to a future study. Surprisingly, the
deuteron $p_T$ spectrum obtained from Eq.~(\ref{eq:coal}) of the
coalescence model agrees better with the data as shown by the dashed
line.

Results from the dynamical model [Eq.~(\ref{eq:coal})] for the
scaled deuteron $v_2^{}$, i.e., $v_{2,d}^{}/2$, are shown by the
solid line in panel (b) of Fig.~\ref{fig:pt_spec deuteron} as a
function of $p_T^{}/2$. Compared to the experimental data shown by
filled circles from the PHENIX Collaboration and by filled squares
from the STAR Collaboration, our model can describe both reasonably
well, particularly those from the PHENIX
Collaboration~\cite{PHENIX07}. The calculated deuteron $v_2^{}$ is,
however, always positive, which is not consistent with the negative
values of deuteron $v_2^{}$ at small $p_T^{}$ seen in the
preliminary STAR data~\cite{STAR07a}. This may again be related to
our neglect of final-state interactions of produced deuterons. If
the nucleon number scaling of deuteron $v_2$ is exact as in the
naive coalescence model, the scaled deuteron elliptic flow
$v_{2,d}^{}(p_T^{}/2)/2$ should be identical to the nucleon elliptic
flow $v_{2,N}^{}(p_T^{})$ [Eq.~(\ref{v2_N})] shown by the dotted
line. Although some deviations are found between the two curves at
large $p_T^{}$ region, our result supports the idea of the nucleon
number scaling of $v_{2,d}^{}$. Furthermore, our result is very
similar to that of the more general coalescence model for deuteron
production (dashed lines). In the considered energy region, the two
approaches give nearly the same $v_{2,d}^{}$. This may indicate that
the momentum conservation in the coalescence model plays a more
important role than the energy conservation that is not respected by
the coalescence model. Therefore, it would be important to clarify
the differences between our dynamical approach and the coalescence
model.

As we have discussed before, even in the more general coalescence
model [Eq.(\ref{eq:coal})], the momenta of the constituent particles
are still nearly aligned to the momentum of the produced particle.
In the case of deuteron, this is because the deuteron wave function
prefers nearly equal momenta for its constituent nucleons, i.e.
$k=0$ in Eq.~(\ref{Hulthen}). Combined with the momentum
conservation, this leads to $\bm{p}_d^{}/2 \approx \bm{p}_1^{}
\approx \bm{p}_2^{}$, but keeping $\bm{p}_d^{} = \bm{p}_1^{} +
\bm{p}_2^{}$. Therefore, although employing the physical deuteron
wave function can relax the momentum constraint $\bm{p}_d^{}/2 =
\bm{p}_1^{} = \bm{p}_2^{}$, the momenta of the constituent nucleons
remain restricted and take only values around the region that
appears in the naive coalescence model, which in turn brings the
$p_T^{}$ dependence of deuteron $v_2^{}$ to scale approximately
according to its constituent nucleon number.

\vspace{0.5cm}
\begin{figure}[t]
\centering
\includegraphics[width=0.6\textwidth,angle=0]{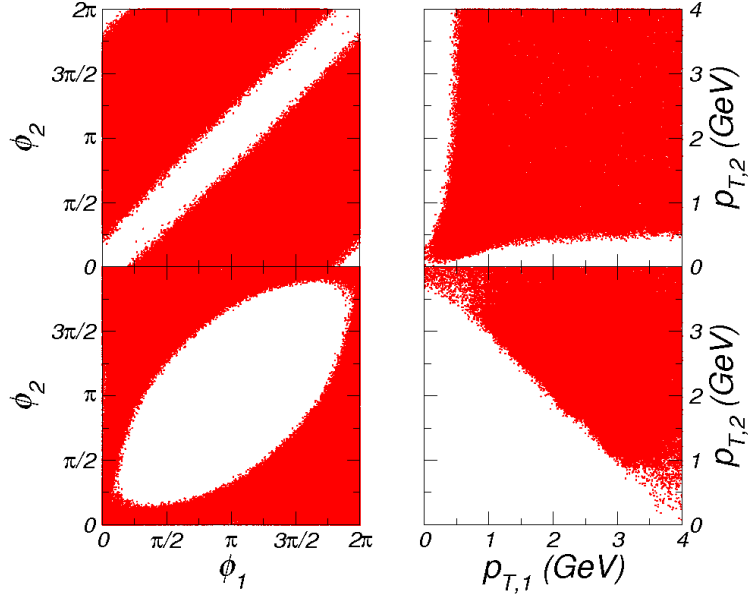}
\caption{(Color online) The allowed regions of $\bm{p}_1^{}$ and
$\bm{p}_2^{}$ in $NN \to d\pi$ reaction. The empty space represents
the unphysical phase space. Left panels show the phase space in the
azimuthal angles $\phi_1$ and $\phi_2$ of the two nucleons and right
panels show the phase space in $|\bm{p}_{T,1}^{}|$ and
$|\bm{p}_{T,2}^{}|$. Upper panels are for $p_{T,d}^{} = 0.1$ GeV and
lower panels for $p_{T,d}^{} = 4.0$ GeV. The deuteron azimuthal
angle is fixed at $\Phi_d = 0$.} \label{fig:kin}
\end{figure}

In the dynamical model, the role of the deuteron wave function
appears through the squared transition amplitude for $NN \to d\pi$,
where the energy is conserved. As a result, the allowed momenta of
the nucleons forming the outgoing deuteron are different from those
in the coalescence model. This can be seen in Fig.~\ref{fig:kin},
which shows the allowed momenta $\bm{p}_{T,1}^{}$ and
$\bm{p}_{T,2}^{}$ of nucleons to form a deuteron of $p_{T,d} = 0.1$
GeV (upper panels) and $4.0$ GeV (lower panels) with $\Phi_d = 0$.
The naive coalescence model allows only one point in each figure,
namely $\phi_1=\phi_2=0$ in the left panels and $p_{T,1}^{} =
p_{T,2}^{}=p_{T,d}^{}/2$ in the right panels. The more general
coalescence model would allow a larger region, but the major
contribution comes from the region of $(p_{T,1}^{},p_{T,2}^{})
\approx (p_{T,d}^{}/2,p_{T,d}^{}/2)$ and $(\phi_1,\phi_2) \approx
(0,0)$ (modulo $2\pi$). The dynamical model based on the $NN \to
d\pi$ reaction covers, on the other hand, the phase space shown by
the shaded areas in the figure, which may or may not include those
involved in the coalescence model. For the phase space in transverse
momenta, because of the exponential form of the nucleon $p_T^{}$
distribution function, the main contribution in the dynamical
approach still comes from that allowed in the more general
coalescence model, i.e. $p_{T,1}^{} \approx p_{T,2}^{}$. For the
phase space in the azimuthal angles, one can find that the region
allowed in the dynamical model for $p_{T,d} = 4.0$ GeV (the
lower-left panel of Fig.~\ref{fig:kin}) covers that of the
coalescence model, which gives the major contribution to deuteron
production. This explains the resemblance of our result with the
coalescence model in the intermediate and large $p_T^{}$ regions.

The similarity between the two models does not hold, however, in the
low $p_T^{}$ region. As shown in upper panels of Fig.~\ref{fig:kin}
for $p_{T,d}^{} = 0.1$ GeV, the kinematically allowed phase space in
the azimuthal angles is very different from the allowed region in
the coalescence model. In fact, the physical phase space excludes
the region of $(\phi_1,\phi_2) \approx (0,0)$ (modulo $2\pi$) as a
result of energy-momentum conservation and the phase space of the
coalescence model is \emph{not} physically allowed. Therefore, the
energy conservation, which is neglected in the coalescence model,
plays an important role in the low $p_T^{}$ region as one would have
expected. In the dynamical model based on the $NN \to d\pi$ reaction
in heavy-ion collisions, there is thus no \emph{a priori} reason to
expect to have the constituent number scaling of the deuteron
$v_2^{}$ at low $p_T^{}$ region.

The scaling behavior of the deuteron $v_2^{}$ shown in the low
$p_T^{}$ region in panel (b) of Fig.~\ref{fig:pt_spec deuteron} is a
coincidence coming from the parameterized nucleon $v_2^{}$ used in
the calculation. In fact, if we allow negative $v_{2,N}^{}$ at very
low $p_T^{}$ region, the coalescence model would predict a negative
$v_{2,d}^{}$ as expected from the scaling behavior. However, in the
present dynamic model calculation, although inclusion of deuteron
final-state interactions is still required, this expectation is not
valid any more in \emph{the low $p_T^{}$ region\/}, and we always
have small but positive $v_{2,d}^{}$ in this region even if we
assume negative nucleon $v_2^{}$ for low $p_T^{}$. This is because
of the fact that the nucleon azimuthal distribution function is
weighted by the nucleon $p_T^{}$ and its distribution function as
can be seen in Eq.~(\ref{eq:det2}). Therefore, for $p_{T,d}^{}=0.1$
GeV, the main contribution to deuteron production comes from
nucleons of $p_{T}^{} \sim 0.5$ GeV, while it must be $\sim 0.05$
GeV in the coalescence model due to the constraint on the magnitude
of the momenta. Furthermore, the physical azimuthal angles in the
dynamical model do not include the region allowed in the coalescence
model. Consequently, these cause the main difference between the
coalescence model and the dynamical model in the present work.

\subsection{Deuteron spectrum and elliptic flow from three-body reactions}

\begin{figure}[t]
\centering
\includegraphics[width=0.6\textwidth,angle=0]{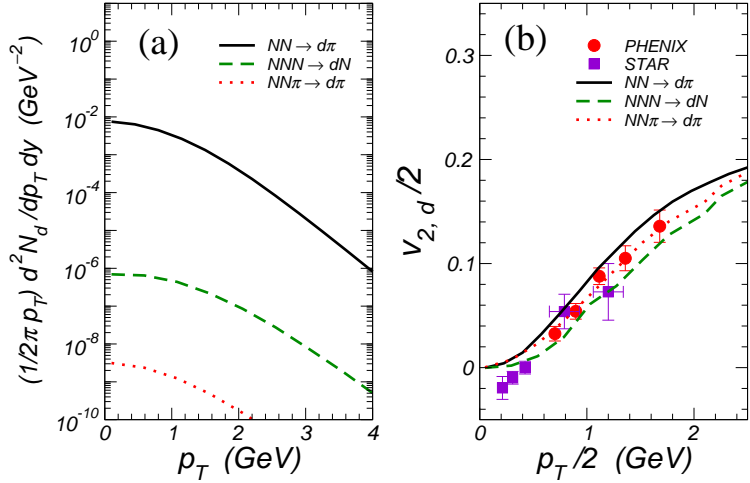}
\caption{(Color online) Calculated deuteron $p_T^{}$ spectrum (panel
(a)) and scaled $v_2^{}$ (panel (b)) at mid-rapidity from $NN\to
d\pi$ (solid lines), $NNN\to dN$ (dashed lines), and $NN\pi\to d\pi$
(dotted lines) reactions. Experimental data are from
Refs.~\cite{PHENIX07,STAR07a}.} \label{fig:v2_d32_EKT}
\end{figure}

For deuteron production from the three-body $NNN \to dN$ and $NN\pi
\to d\pi$ reactions, the deuteron transverse momentum spectrum and
differential elliptic flow can be similarly calculated as in the
case of the two-body $NN\to\ d\pi$ reaction. Multiplying the
calculated rates by the same hadronic matter volume $V\approx
3,700~{\rm fm}^3$ at freeze out and the time interval
$\Delta\tau\approx 4 \mbox{ fm}/c$, the results are shown in
Fig.~\ref{fig:v2_d32_EKT} by the dashed and dotted lines,
respectively. For comparison, results based on the two-body $NN \to
d\pi$ reaction, which have been shown in Fig.~\ref{fig:pt_spec
deuteron}, are also given by solid lines.

Although the physically allowed nucleon phase spaces for deuteron
production in $NNN$ and $NN\pi$ reactions are different from that in
the $NN \to d\pi$ reaction, the predicted deuteron $v_2^{}$ from the
three reactions do not show drastic difference from each other as
shown in panel (b) of Fig.~\ref{fig:v2_d32_EKT}, and all lie close
to the experimental data except at very low momentum where the
deuteron $v_2$ is negative in the preliminary data from the STAR
Collaboration~\cite{STAR07a}.

In contrast to the deuteron $v_2^{}$, the predicted deuteron
transverse momentum spectrum depends strongly on the reaction
mechanism. As shown in panel (a) of Fig.~\ref{fig:v2_d32_EKT},
contributions from three-body reactions are suppressed by several
orders of magnitude compared to that from the two-body reaction.
This is mainly due to the low nucleon and pion densities at freeze
out. Another possible, but subsidiary, reason for this behavior is
the magnitude of the reaction amplitudes used in our study. The very
simple model we have used for the $NNN \to dN$ and $NN\pi \to d\pi$
reactions may have underestimated their contributions to deuteron
production. On the contrary to the phase space in $NN \to d\pi$
reaction (Fig.~\ref{fig:kin}), the phase space in the
three-to-two-body reactions has no restrictions on the angles and
magnitudes of the momenta of three initial particles.

\section{Summary and discussions}
\label{summary}

In this paper, we have studied the deuteron elliptic flow in
heavy-ion collisions at RHIC using a dynamical model for deuteron
production that conserves both momentum and energy. Since elliptic
flows in relativistic heavy ion collisions are mainly generated
during early partonic stage, deuteron elliptic flow is expected to
not change much during the hadronic stage, we have considered the
deuteron production rate from the $NN \to d\pi$ reaction at the
thermal freeze out. The deuteron yield is then obtained from
multiplying the production rate by the volume of the hadronic matter
at freeze out and a time interval of about 4~fm/$c$. Because of the
neglect of deuteron final-state interactions, which would enhance
the incomplete treatment of radial flow effect, the resulting
deuteron $p_T^{}$ spectrum is somewhat softer than that measured in
experiments. The obtained result for the deuteron $v_2^{}$ agrees,
however, reasonably with the measured one, except at very low
$p_T^{}$ where the latter is negative. Our results further show that
this model can reproduce the predictions from the coalescence model.
The origin of this result could be understood from the physically
allowed phase space of nucleon momenta in deuteron production. At
large $p_T^{}$, the phase space in our dynamical model covers that
of the coalescence model, which eventually dominate the production
process because of the nucleon distribution functions. Therefore,
our dynamical model supports the description of the coalescence
model. However, at very low $p_T^{}$, the nucleon phase space in the
coalescence model turns out to be physically forbidden because of
energy-momentum conservation. Thus, deuteron production via $NN \to
d\pi$ generally does not reproduce the prediction of the coalescence
model at very low $p_T^{}$. Although the prediction from the
dynamical model follows the scaling behavior, this property is
strongly dependent of the nucleon $v_2^{}$ at the low $p_T^{}$
region.

We have also considered three-body $NNN \to dN$ and $NN\pi \to d\pi$
reactions for deuteron production in heavy-ion collisions. Because
of the different nucleon phase space involved in these reactions
from that in the two-body $NN\to d\pi$ reaction, these reactions
lead to different predictions on the deuteron $v_2^{}$. The deuteron
production rates from these reactions are, however, suppressed
compared to that from the two-body $NN \to d\pi$ reaction by several
orders of magnitude. This implies that deuteron production in
heavy-ion collisions is mainly due to the $NN \to d\pi$ reaction,
which has the largest cross section among the $NN \to d X$
reactions~\cite{ADEK71}. We have also found that all models
considered in the present study give positive deuteron $v_2^{}$ at
very low $p_T^{}$. This is in contrast to preliminary data from the
STAR Collaboration that deuterons produced in heavy ion collisions
at RHIC may have negative $v_{2}^{}$ at very low $p_T^{}$ region. To
obtain negative deuteron $v_2$ required negative nucleon $v_2$ in
the coalescence model. Therefore, precise measurements of both the
nucleon and deuteron $v_2^{}$ in this transverse momentum region are
highly desirable for understanding the deuteron transverse momentum
and elliptic flow in heavy-ion collisions. For our dynamical model
approach, there are several issues which deserve further studies. As
mentioned in Section~\ref{results}, final-state interactions of
deuterons, particularly the $\pi d$ elastic scattering, need to be
taken into account via a microscopic transport model or a schematic
kinetic model. Also, the deuteron production amplitude used in the
present study assumes a point-like structure of the deuteron. For a
more realistic description, the spread of the deuteron wave function
should be taken into account in the production amplitude to render a
more direct comparison with the coalescence model.

\acknowledgments

We are grateful to Wei Liu for useful discussions. This work was
supported by the US National Science Foundation under Grant No.\
PHY-0457265 and the Welch Foundation under Grant No.\ A-1358.

\appendix

\section{\boldmath $NN \to d\pi$ reaction}\label{two}

\begin{figure}[t]
\centering
\includegraphics[width=0.5\textwidth,angle=0]{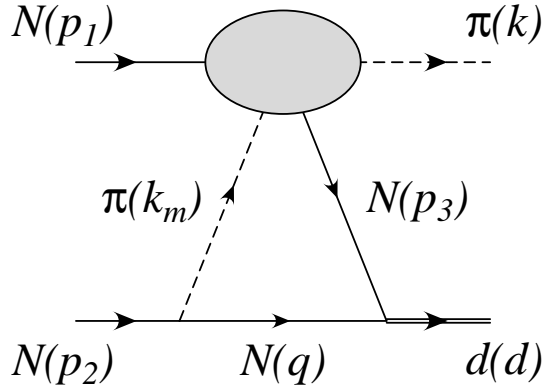}
\caption{Diagram for the $NN \to d\pi$ reaction. The blob denotes
the $\pi N \to \pi N$ elastic scattering.} \label{fig:nn_dpi}
\end{figure}

The $NN \to d\pi$ reaction can be described by the one-pion exchange
diagram shown in Fig.~\ref{fig:nn_dpi}. In this model, one of the
incoming nucleons emits a virtual pion which causes the $\pi N \to
\pi N$ reaction shown by the blob in Fig.~\ref{fig:nn_dpi} before
forming the outgoing deuteron. There are other diagrams which
include the nucleon exchange diagrams but it is well-known that
their contributions are negligible~\cite{GKKLS84}. In this study, we
thus adopt a simple model for the $NN \to d\pi$ reaction of
Fig.~\ref{fig:nn_dpi} following the method of
Refs.~\cite{Yao64,Barry72}. Here we focus on the reaction of $pp \to
d\pi^+$ and use the isospin relations $\sigma(pp \to d\pi^+) =
\sigma(nn \to d\pi^-) = 2\sigma(pn \to d\pi^0)$ to obtain the
isospin averaged cross section $\sigma(NN\to d\pi)=\sigma(pp\to
d\pi^+)$.

Using the momenta indicated in Fig.~\ref{fig:nn_dpi}, the amplitude
for the $pp \to d\pi^+$ reaction can be written as
\begin{eqnarray}
T &=& - \frac{i}{(2\pi)^4} \int d^4 q \bar{v}(p_2^{}) \Gamma_{\pi
NN} \frac{1}{(p_2-q)^2 - M_\pi^2} \frac{-q \cdot \gamma +
M_N}{q^2-M_N^2} \Gamma_{dNN} \nonumber \\ && \qquad \mbox{} \times
\frac{(d-q)\cdot\gamma + M_N}{(d-q)^2 - M_N^2} \left( A + B
k\cdot\gamma \right) u(p_1),
\end{eqnarray}
where $M_N$ and $M_\pi$ are, respectively, the nucleon
and pion masses, and $\Gamma_{\pi NN}$ and $\Gamma_{dNN}$
denote, respectively, the pion-nucleon and deuteron-nucleon
vertices. The functions $A$ and $B$ depend on the Mandelstam
variables and are related to the invariant $\pi N$ elastic
scattering amplitude by
\begin{equation}
\mathcal{M}( \pi N \to \pi N) = \bar{u}(p_f^{}) \left( A + B k
\cdot\gamma \right) u(p_i^{}),
\end{equation}
where $p_i^{}$ and $p_f^{}$ are the initial and final nucleon
momenta, respectively. The squared matrix element can be written in
terms of the $\pi N$ elastic scattering cross section as
\begin{equation}
|\mathcal{M}(\pi N \to \pi N)|^2 = (8\pi)^2 2 s_{\pi N}^{}
\frac{d\sigma_{\pi N}^{}}{d\Omega_{\pi N}^{}}(s_{\pi N}^{}, u_{\pi
N}^{}).
\end{equation}
where the Mandelstam variables for the $\pi N$ system are $s_{\pi
N}^{} = (k_i^{} + p_i^{})^2$ and $u_{\pi N}^{} = (k_i^{}-p_f^{})^2$,
with $k_i^{}$ being the momentum of the initial pion. In our
qualitative study, the elastic $\pi N$ scattering can be well
approximated by assuming that it is dominated by the $\Delta$
resonance with the interaction Lagrangian
\begin{equation}
\mathcal{L}_{\pi N\Delta} = \frac{f_{\pi N\Delta}}{M_\pi}
\bar{\Delta}^\mu \partial_\mu \pi N + \mbox{ H.c.},
\end{equation}
where $f_{\pi N\Delta} = 2.23$ and the isospin factors can be found,
e.g., in Ref.~\cite{ONL04}.

\begin{figure}[tb]
\centering
\includegraphics[width=0.5\textwidth,angle=0]{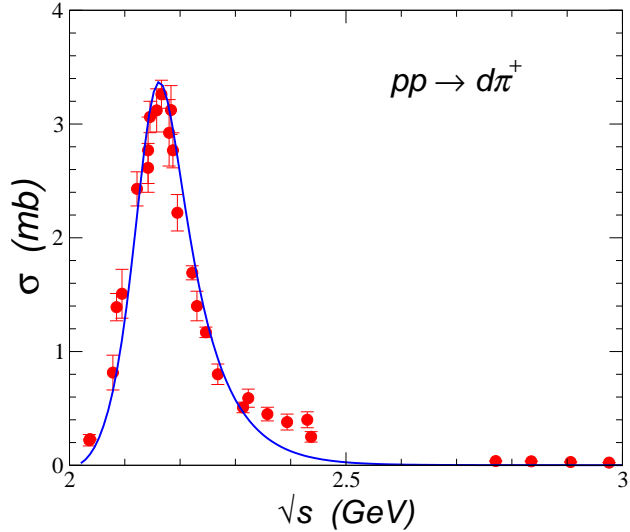}
\caption{(Color online) Total cross section for the $pp \to d\pi^+$
reaction. The experimental data are from
Refs.~\cite{ALMD74,SKKSSY82,GEM00,COSY06}.} \label{fig:nn_dpi_sigma}
\end{figure}

Following Ref.\cite{GKKLS84}, the spin averaged squared amplitude is
then given by
\begin{eqnarray}
|\mathcal{M}|^2 &=& \frac14 \sum_{\rm spins} |T|^2 \nonumber \\
&=& -\left[ \frac{\kappa}{M_d} \frac{g_{\pi NN}G_{\pi
NN}(k_m^2)}{k_m^2-M_\pi^2} \right]^2 \left[ \frac{g_{dNN}^{}}{M_N}
\frac{1}{2\sqrt2 M_d} \right]^2 \frac12 k_m^2 M_d^2 (2 M_N + M_d)^2
\nonumber \\ && \mbox{} \times \left\{3 \left( 1 +
\frac{\varrho}{\sqrt2} \right)^2 - \frac{6\varrho}{\sqrt2} \left( 1
+ \frac{\varrho}{\sqrt2} \right) + \left( \frac{3\varrho}{\sqrt2}
\right)^2 \right\} |\mathcal{M}(\pi N \to \pi N)|^2. \label{amp1}
\end{eqnarray}
In the above, $\varrho$ $(\approx 4\%)$ is the amount of $D$-wave
state in the deuteron wave function, $M_d$ is the deuteron mass, and
$\kappa \equiv \sqrt{M_N\varepsilon}$ with the deuteron binding
energy $\varepsilon$ ($\approx 2.23$ MeV). The pion-nucleon coupling
constant is denoted by $g_{\pi NN}$ with a value $g^2_{\pi NN}/4\pi
= 14.0$, and $G_{\pi NN}(q^2)$ is the form factor
\begin{equation}
G_{\pi NN}(q^2) = \frac{\Lambda^2 -M_\pi^2}{\Lambda^2 - q^2}
\end{equation}
with $\Lambda = 0.73$ GeV. The deuteron-nucleon coupling constant
$g_{dNN}^{}$ is given by
\begin{equation}
g_{dNN}^{} = \left[
\frac{16\pi\kappa}{M_N(1-r_t^{}\kappa)(1+\varrho^2)} \right]^{1/2},
\end{equation}
where $r_t^{}$ ($\approx 1.75$ fm) is the effective range for the
spin-triplet nucleon-nucleon elastic scattering. To reproduce the
experimental total cross section for $pp \to d\pi^+$ within this
simple model, we use $M_\Delta = 1200$ MeV. The calculated cross
section is shown in Fig.~\ref{fig:nn_dpi_sigma} and is seen to
reproduce very well the measured one.

\section{\boldmath $NNN\to dN$ and $NN\pi\to d\pi$
reactions}\label{three}

The amplitudes for producing deuterons from three-body reactions
$NNN\to dN$ and $NN\pi\to d\pi$ are approximated by making use of
the amplitude of $NN \to d\pi$ obtained above. For the $NNN \to dN$
reaction, we consider the one-pion exchange process, namely, the
collision of two initial nucleons gives the deuteron and an
off-shell pion, which is then absorbed by the spectator nucleon. In
this approximation, the reaction amplitude becomes
\begin{equation}
|\mathcal{M}(NNN \to dN)|^2 = \frac{4G_{\pi NN}^2}{(q_m^2 - M_\pi^2)^2}
\left( p_3^{} \cdot p_f^{} - M_N^2 \right) | \mathcal{M}(NN \to d\pi)|^2,
\end{equation}
where $q_m^{}$ is the intermediate pion momentum, $p_3^{}$ is the
momentum of the initial spectator nucleon, and $p_f^{}$ is the
nucleon momentum in the final state.

\begin{figure}[tb]
\centering
\includegraphics[width=0.6\textwidth,angle=0]{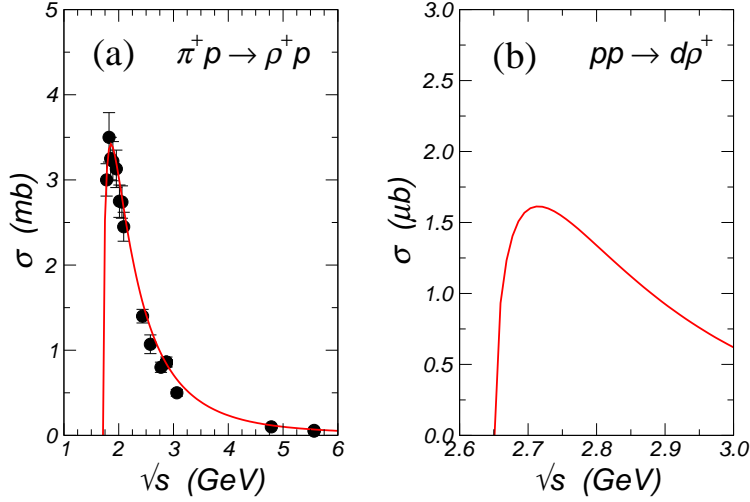}
\caption{(Color online) Total cross sections for $\pi^+ p \to \rho^+
p$ (panel (a) and $pp \to d\rho^+$ (panel (b)). Experimental data
are from Ref.~\cite{WFKM72}.}
\label{fig:pi_rho}
\end{figure}

Replacing the off-shell pion by an off-shell rho-meson, which is
then absorbed by a spectator pion leads to an approximate
description of the $NN\pi \to d\pi$ reaction. The amplitude for the
subprocess $NN \to d\rho$ can be obtained by the same method of
previous section for obtaining the $NN \to d\pi$ amplitude. The
result is thus similar to Eq.~(\ref{amp1}) with $|\mathcal{M}(\pi N
\to \pi N)|^2$ replaced by $|\mathcal{M}(\pi N \to \rho N)|^2/2$. A
complete analysis of the $\pi N \to \rho N$ reaction requires many
possible terms arising from meson exchanges and nucleon resonance
contributions. For a simple estimate on the cross section for this
reaction, we describe it by one-pion exchange only and fit the
experimental data with suitable form factors. The relevant effective
Lagrangians are
\begin{eqnarray}
\mathcal{L}_{\rho\pi\pi} &=& g_{\rho\pi\pi}^{} \bm{\rho}^\mu \cdot
\left( \bm{\pi} \times \partial_\mu \bm{\pi} \right), \nonumber \\
\mathcal{L}_{\pi NN} &=& - \frac{g_{\pi NN}}{2M_N} \bar{N} \bm{\tau}
\cdot \partial_\mu \bm{\pi} \gamma_5 \gamma^\mu N,
\end{eqnarray}
where $g_{\rho\pi\pi}^{} = 6.02$ and $g_{\pi NN}^2/(4\pi) = 14.0$.
The form factor in the form of $\Lambda^4/[\Lambda^4 +
(t-M_\pi^2)^2]$ is used with $\Lambda = 0.65$ GeV. The $NN\pi \to
d\pi$ amplitude is then given by
\begin{equation}
|\mathcal{M}(NN\pi \to d\pi)|^2 = \frac{2G_{\rho NN}^2}{(q_m^2 - M_\rho^2)^2}
\left( p_3^{} \cdot p_f^{} - M_\pi^2 \right) | \mathcal{M}(NN \to d\rho)|^2.
\end{equation}
Our results on the total cross sections for $\pi^+ p \to \rho^+ p$
and $pp \to d \rho^+$ are shown in Fig.~\ref{fig:pi_rho}. The former
is seen to agree with the measured cross section.


\begin{thebibliography}{10}

\bibitem{Olli92}
J.-Y. Ollitrault,
\newblock Phys.\ Rev.\ D {\bf 46}, 229 (1992).
%%CITATION = PHRVA,D46,229;%%

\bibitem{VZ94}
S.~Voloshin and Y.~Zhang,
\newblock Z.\ Phys.\ C {\bf 70}, 665 (1996)
[arXiv:hep-ph/9407282].
%%CITATION = HEP-PH 9407282;%%

\bibitem{RR97}
W.~Reisdorf and H.~G. Ritter,
\newblock Ann.\ Rev.\ Nucl.\ Part.\ Sci.\ {\bf 47}, 663 (1997).
%%CITATION = ARNUA,47,663;%%

\bibitem{DLGP98}
P.~Danielewicz, R.~A.~Lacey, P.~B.~Gossiaux, C.~Pinkenburg,
P.~Chung, J.~M.~Alexander and R.~L.~McGrath,
\newblock Phys.\ Rev.\ Lett.\ \textbf{81}, 2438 (1998)
[arXiv:nucl-th/9803047].
%%CITATION = NUCL-TH 9803047;%%

\bibitem{Lacey05}
R.~A. Lacey,
\newblock Nucl.\ Phys.\ A {\bf 774}, 199 (2006)
[arXiv:nucl-ex/0510029].
%%CITATION = NUCL-EX 0510029;%%

\bibitem{STAR05c}
STAR Collaboration, J.~Adams {\em et~al.},
\newblock Nucl.\ Phys.\ A {\bf 757}, 102 (2005)
  [arXiv:nucl-ex/0510009].
%%CITATION = NUCL-EX 0501009;%%

\bibitem{PHENIX05a}
PHENIX Collaboration, K.~Adcox {\em et~al.},
\newblock Nucl.\ Phys.\ A {\bf 757}, 184 (2005)
[arXiv:nucl-ex/0410003].
%%CITATION = NUCL-EX 0410003;%%

\bibitem{KSH00}
P.~F. Kolb, J.~Sollfrank, and U.~Heinz,
\newblock Phys.\ Rev.\ C {\bf 62}, 054909 (2000)
[arXiv:hep-ph/0006129].
%%CITATION = HEP-PH 0006129;%%

\bibitem{LK01}
Z.~W.~Lin and C.~M.~Ko,
Phys.\ Rev.\ C {\bf 65}, 034904 (2002)
[arXiv:nucl-th/0108039].
%%CITATION = NUCL-TH 0108039;%%

\bibitem{GKL03-GKL03b}
V.~Greco, C.~M. Ko, and P.~L\'{e}vai,
Phys.\ Rev.\  Lett.\ \textbf{90}, 202302 (2003)
[arXiv:nucl-th/0301093];
Phys. Rev. C \textbf{68}, 034904 (2003)
[arXiv:nucl-th/0305024].
%%CITATION = NUCL-TH 0301093;%%
%%CITATION = NUCL-TH 0305024;%%

\bibitem{HY02-HY03}
R.~C. Hwa and C.~B. Yang,
Phys.\ Rev.\ C \textbf{67}, 034902 (2003)
[arXiv:nucl-th/0211010];
Phys.\ Rev.\ C \textbf{67}, 064902 (2003)
[arXiv:nucl-th/0302006].
%%CITATION = NUCL-TH 0211010;%%
%%CITATION = NUCL-TH 0302006;%%

\bibitem{FMNB03-FMNB03b}
R.~J. Fries, B.~M\"{u}ller, C.~Nonaka, and S.~A. Bass,
Phys.\ Rev.\ Lett.\ \textbf{90}, 202303 (2003)
[arXiv:nucl-th/0301087];
Phys.\ Rev.\ C \textbf{68}, 044902 (2003)
[arXiv:nucl-th/0306027].
%%CITATION = NUCL-TH 0301087;%%
%%CITATION = NUCL-TH 0306027;%%

\bibitem{MV03}
D.~Moln{\'a}r and S.~A. Voloshin,
\newblock Phys.\ Rev.\ Lett.\ {\bf 91}, 092301 (2003)
[arXiv:nucl-th/0302014].
%%CITATION = NUCL-TH 0302014;%%

\bibitem{KCGK04}
P.~F. Kolb, L.-W. Chen, V.~Greco, and C.~M. Ko,
\newblock Phys. Rev. C {\bf 69}, 051901(R) (2004)
[arXiv:nucl-th/0402049].
%%CITATION = NUCL-TH 0402049;%%

\bibitem{PHENIX06}
PHENIX Collaboration, S.~Adare {\em et~al.},
\newblock Phys. Rev. Lett. {\bf 98}, 162301 (2007)
[arXiv:nucl-ex/0608033].
%%CITATION = NUCL-EX 0608033;%%

\bibitem{PHENIX07}
PHENIX Collaboration, S.~Afanasiev {\em et~al.},
\newblock Phys. Rev. Lett. {\bf 99}, 052301 (2007)
[arXiv:nucl-ex/0703024].
%%CITATION = NUCL-EX 0703024;%%

\bibitem{STAR07a}
STAR Collaboration, H.~Liu,
\newblock J. Phys. G {\bf 34}, S1087 (2007)
[arXiv:nucl-ex/0701057].
%%CITATION = NUCL-EX 0701057;%%

\bibitem{DHSZ91}
C.~B. Dover, U.~Heinz, E.~Schnedermann, and J.~Zim{\'a}nyi,
\newblock Phys. Rev. C {\bf 44}, 1636 (1991).
%%CITATION = PHRVA,C44,1636;%%

\bibitem{McL07}
L.~McLerran,
\newblock arXiv:hep-ph/0702004.
%%CITATION = HEP-PH 0702004;%%

\bibitem{RR07}
L.~Ravagli and R.~Rapp,
\newblock arXiv:0705.0021.
%%CITATION = ARXIV:0705.0021;%%

\bibitem{LKLZP04}
Z.~W.~Lin, C.~M.~Ko, B.~A.~Li, B.~Zhang, and S.~Pal,
\newblock Phys.\ Rev.\ C {\bf 72}, 064901 (2005)
[arXiv:nucl-th/0411110].
%%CITATION = NUCL-TH 0411110;%%

\bibitem{CGKLL03}
L.~W. Chen, V.~Greco, C.~M. Ko, S.~H. Lee, and W.~Liu,
\newblock Phys.\ Lett.\ B {\bf 601}, 34 (2004)
[arXiv:nucl-th/0308006].
%%CITATION = NUCL-TH 0308006;%%

\bibitem{ADEK71}
H.~L. Anderson {\em et~al.\/},
\newblock Phys.\ Rev.\ Lett.\ {\bf 26}, 108 (1971).
%%CITATION = PRLTA,26,108;%%

\bibitem{Mattie95}
R. Mattiello, A. Jahns, H. Sorge, H. St{\"o}cker, and W. Greiner,
\newblock Phys.\ Rev.\ Lett.\ \textbf{74}, 2180 (1995);
%%CITATION = PRLTA,74,2180;%%
R. Mattiello, H. Sorge, H. St{\"o}cker, and W. Greiner,
\newblock Phys.\ Rev.\ C \textbf{55}, 1443 (1997)
[arXiv:nucl-th/9607003].
%%CITATION = NUCL-TH 9607003;%%

\bibitem{CKL03}
L.-W. Chen, C.~M. Ko, and B.~A. Li,
Phys. Rev. C {\bf 68}, 017601 (2003)
[arXiv:nucl-th/0302068].
%%CITATION = NUCL-TH 0302068;%%

\bibitem{CKL03a}
L.-W. Chen, C.~M. Ko, and B.-A. Li,
\newblock Nucl.\ Phys.\ A {\bf 729}, 809 (2003)
[arXiv:nucl-th/0306032].
%%CITATION = NUCL-TH 0306032;%%

\bibitem{yan}
T.~Z. Yan {\it et~al.\/},
Phys. Lett. B {\bf 638}, 50 (2006)
[arXiv:nucl-th/0605022].
%%CITATION = NUCL-TH 0605022;%%

\bibitem{CK86}
L.~P. Csernai and J.~I. Kapusta,
\newblock Phys.\ Rep.\ {\bf 131}, 223 (1986).
%%CITATION = PRPLC,131,223;%%

\bibitem{DB91}
P.~Danielewicz and G.~F. Bertsch,
\newblock Nucl.\ Phys.\ A {\bf 533}, 712 (1991).
%%CITATION = NUPHA,A533,712;%%

\bibitem{DS93}
A.~Deloff and T.~Siemiarczuk,
\newblock Nucl.\ Phys.\ A {\bf 555}, 659 (1993).
%%CITATION = NUPHA,A555,659;%%

\bibitem{GSTH06}
D.~E. Gonzalez~Trotter {\em et~al.},
\newblock Phys.\ Rev.\ C {\bf 73}, 034001 (2006).
%%CITATION = PHRVA,C73,034001;%%

\bibitem{PHENIX03b}
PHENIX Collaboration, S.~S. Adler {\em et~al.},
\newblock Phys.\ Rev.\ C {\bf 69}, 034909 (2004)
[arXiv:nucl-ex/0307022].
%%CITATION = NUCL-EX 0307022;%%

\bibitem{PHENIX05}
PHENIX Collaboration, S.~S. Adler {\em et~al.},
\newblock Phys.\ Rev.\ Lett.\ {\bf 94}, 122302 (2005)
[arXiv:nucl-ex/0406004].
%%CITATION = NUCL-EX 0406004;%%

\bibitem{Gumbel}
R.~A. Fisher and L.~H.~C. Tippett,
Proc.\ Cambridge Phil.\ Soc.\ {\bf 24}, 180 (1928);
E.~J. Gumbel,
{\it Statistics of Extremes},
(Columbia University Press, New York, 1958).
%%CITATION = PCPSA,24,180;%%

\bibitem{GK04a}
V.~Greco and C.~M. Ko,
Phys.\ Rev.\ C {\bf 70}, 024901 (2004)
[arXiv:nucl-th/0402020].
%%CITATION = NUCL-TH 0402020;%%

\bibitem{KHHH00}
P.~F. Kolb, P.~Huovinen, U.~Heinz, and H.~Heiselberg,
\newblock Phys.\ Lett.\ B {\bf 500}, 232 (2001)
[arXiv:hep-ph/0012137].
%%CITATION = HEP-PH 0012137;%%

\bibitem{HKHRV01}
P.~Huovinen {\em et~al.},
\newblock Phys.\ Lett.\ B {\bf 503}, 58 (2001)
[arXiv:hep-ph/0101136].
%%CITATION = HEP-PH 0101136;%%

\bibitem{DESXX04}
X.~Dong {\em et~al.},
\newblock Phys.\ Lett.\ B {\bf 597}, 328 (2004)
[arXiv:nucl-th/0403030].
%%CITATION = NUCL-TH 0403030;%%

\bibitem{KDHM91}
M.~A. Khandaker, M.~Doss, I.~Halpern, T.~Murakami, D.~W. Storm,
D.~R. Tieger, and W.~J. Burger,
\newblock Phys.\ Rev.\ C {\bf 44}, 24 (1991).
%%CITATION = PHRVA,C44,24;%%

\bibitem{art}
B.~A. Li and C.~M. Ko,
\newblock Phys.\ Rev.\ C {\bf 52}, 2037 (1995)
[arXiv:nucl-th/9505016];
%%CITATION = NUCL-TH 9505016;%%
B.~A. Li, A.~T. Sustich, B. Zhang, and C.~M. Ko,
\newblock Int.\ Jour.\ Mod.\ Phys.\ E {\bf 10}, 267 (2001).
%%CITATION = IMPAE,E10,267;%%

\bibitem{GKKLS84}
W.~Grein {\em et~al.},
\newblock Ann.\ Phys.\ (N.Y.) {\bf 153}, 301 (1984).
%%CITATION = APNYA,153,301;%%

\bibitem{Yao64}
T.~Yao,
\newblock Phys.\ Rev.\ {\bf 134}, B454 (1964).
%%CITATION = PHRVA,134,B454;%%

\bibitem{Barry72}
G.~W. Barry,
\newblock Ann.\ Phys.\ (N.Y.) {\bf 73}, 482 (1972).
%%CITATION = APNYA,73,482;%%

\bibitem{ONL04}
Y.~Oh, K.~Nakayama, and T.-S.~H. Lee,
\newblock Phys.\ Rep.\ {\bf 423}, 49 (2006)
[arXiv:hep-ph/0412363].
%%CITATION = HEP-PH 0412363;%%

\bibitem{ALMD74}
H.~L. Anderson {\em et~al.},
\newblock Phys.\ Rev.\ D {\bf 9}, 580 (1974).
%%CITATION = PHRVA,D9,580;%%

\bibitem{SKKSSY82}
F.~Shimizu {\em et~al.},
\newblock Nucl.\ Phys.\ A {\bf 386}, 571 (1982).
%%CITATION = NUPHA,A386,571;%%

\bibitem{GEM00}
GEM Collaboration, M.~Betigeri {\em et~al.},
\newblock Phys.\ Rev.\ C {\bf 63}, 044011 (2001)
[arXiv:nucl-ex/0101001].
%%CITATION = NUCL-EX 0101001;%%

\bibitem{COSY06}
COSY-TOF Collaboration, S.~Abd El-Samad {\em et~al.},
\newblock Eur.\ Phys.\ J.\ A {\bf 30}, 443 (2006)
[arXiv:nucl-ex/0609011].
%%CITATION = NUCL-EX 0609011;%%

\bibitem{WFKM72}
Y.~Williamson {\em et~al.},
\newblock Phys.\ Rev.\ Lett.\ {\bf 29}, 1353 (1972);
%%CITATION = PRLTA,29,1353;%%
Reaction Data Database, http://durpdg.dur.ac.uk/HEPDATA/REAC.

\end{thebibliography}
\end{document}